\documentclass[12pt]{article}
\usepackage{amsmath}
\usepackage[left=1in,top=1in,right=1in,nohead]{geometry}
\usepackage{graphics,graphicx,epsfig}
\usepackage{amssymb}
\usepackage{epstopdf}
\usepackage{color}
\usepackage{sidecap}
\usepackage{appendix}
\usepackage{amsmath}

\def\calm{{\cal M}}
\def\calt{{\cal T}}

\def\obar{  \bar{\Omega}_* }

\providecommand{\ev}[1]{\left\langle{#1}\right\rangle}

\newcommand{\Tr}{\mathrm{Tr}}

\def\be#1\ee{\begin{equation}#1\end{equation}}
\def\bea#1\eea{\begin{align}#1\end{align}}

\begin{document}

\begin{titlepage}
\vfill
\begin{flushright}
ACFI-T14-13
\end{flushright}

\vfill
\begin{center}
\baselineskip=16pt
{\Large\bf Non-Vacuum AdS Cosmologies and the Approach to Equilibrium of Entanglement Entropy}


\vskip 0.2in
{\large {\sl }}
\vskip 10.mm
{\large\bf 
Sebastian Fischetti$^{a,}$\footnote{sfischet@physics.ucsb.edu} David Kastor$^{b,}$\footnote{kastor@physics.umass.edu} and Jennie Traschen$^{b,}$\footnote{traschen@physics.umass.edu}} 

\vskip 0.5cm
{{$^a$Department of Physics, University of California, Santa Barbara, CA 93106\\
$^{b}$Amherst Center for Fundamental Interactions, Department of Physics\\ University of Massachusetts, Amherst, MA 01003
     }}
\vspace{6pt}
\end{center}
\vskip 0.2in
\par
\begin{center}
{\bf Abstract}
 \end{center}
\begin{quote}
We extend standard results for vacuum asymptotically locally AdS (AlAdS) spacetimes, showing that such spacetimes can be constructed as foliations where the induced metric on each hypersurface satisfies Einstein's equation with stress-energy.  By an appropriate choice of stress-energy on the hypersurfaces, the resulting AlAdS spacetime satisfies Einstein's equation with a negative cosmological constant and physical stress tensor.  We use this construction to obtain AlAdS solutions whose boundaries are FRW cosmologies sourced by a massless scalar field or by a perfect fluid obeying the strong energy condition.  We focus on FRW universes that approach Minkowski spacetime at late times, yielding AlAdS spacetimes that approach either the Poincar\'e patch of pure AdS or the AdS soliton, which we view as late time equilibrium states.  As an application of these solutions, we use the AdS/CFT correspondence to study the approach to equilibrium of the entanglement entropy and of the boundary stress tensor of the boundary CFT.  We find that the energy of the asymptotically AdS solitonic solution is consistent with the conjecture that the AdS soliton is the lowest-energy solution to Einstein's equation with negative cosmological constant.  The time dependent correction to the entanglement entropy is found to decay like a power law, with rate set by the Hubble parameter and the power determined by the equation of state of the cosmic fluid.



  \vfill
\vskip 2.mm
\end{quote}
\hfill
\end{titlepage}


\tableofcontents

\section{Introduction}

A large class of interesting and useful asymptotically locally anti-de Sitter (AlAdS) spacetimes have been constructed by starting with AdS in Poincar\'e coordinates, in which the spacetime is foliated by slices on which the metric is conformal to the Minkowski metric $\eta_{ab}$, and replacing $\eta_{ab}$ with any Ricci-flat metric $\gamma_{ab}$.  Thus~$(D-1)$-dimensional vacuum solutions to Einstein's equations straightforwardly give rise to new $D$-dimensional solutions to Einstein's equation with a negative cosmological constant.  For example, taking $\gamma_{ab}$ to be the Schwarzchild black hole yields a black cigar \cite{Chamblin:1999by}, and $\gamma_{ab}$ was taken to be a vacuum $pp$-wave in \cite{Chamblin:1999cj} to construct a wave  in the far field of an AdS-brane spacetime.  Furthermore, the AlAdS solutions so generated are of particular interest in light of the AdS/CFT correspondence~\cite{Maldacena:1997re,Gubser:1998bc,Witten:1998qj}, since they are dual to a large-$N$, strongly coupled conformal field theory (CFT) that lives on the spacetime~$\gamma_{ab}$ (or a conformally rescaled version thereof).  For instance, \cite{Engelhardt:2013jda,Engelhardt:2014mea} took~$\gamma_{ab}$ to be a vacuum Kasner metric in order to study a cosmological singularity by computing the entanglement entropy and Wightman functions of the CFT\footnote{In fact, in~\cite{Engelhardt:2014mea} the CFT lived on a singularity-free conformally rescaled version of Kasner.}.

In this paper we generalize this construction and show how non-vacuum $(D-1)$-dimensional spacetimes can be used to give AlAdS spacetimes with nonzero stress-energy. Specifically, if $\gamma_{ab}$ is a solution to the Einstein equations in $(D-1)$-dimensions with stress-energy tensor $\widehat{T}_{ab}$, we show that replacing $\eta_{ab}$ on the Poincar\'e slices with $\gamma_{ab}$ gives a $D$-dimensional AdS solution with stress-energy $T_{ab}$ that satisfies
\be
T_{\mu\nu} = \widehat{T}_{\mu\nu},
\ee
where~$\mu,\nu = 0,\ldots,D-1$.  By appropriate choice of~$\widehat{T}_{ab}$, we use this construction to find new AlAdS spacetimes that have a physically sensible stress-energy.  We also show that these spacetimes can be ``solitonized'' \cite{Haehl:2012tw} by adding a compact dimension that shrinks smoothly to zero in the AdS bulk.  A variation of this non-vacuum construction was performed in \cite{Cvetic:2000gj,Lu:2000xc,Park:2001jh}, which studied getting  supergravity gauge fields ``on the brane'' by doing a Kaluza-Klein reduction of a supergravity theory in the higher dimensional AdS spacetime. A related
construction starting with supergravity fields in ten dimensions was used to explore properties of time dependent boundaries in references
 \cite{Das:2006dz,Das:2006pw,Awad:2007fj,Awad:2008jf}.
 The relation between their higher and lower dimensional matter theories differs from what we find here, as will be clarified in Section~\ref{derive}. 

As a special example, we will take~$\gamma_{ab}$ to be a Friedman-Robertson-Walker (FRW) cosmology\footnote{From the CFT side, such solutions can be thought of as a generalization of those in~\cite{Koyama:2001rf}, which took the boundary metric to be a conformally flat FRW geometry.}.  The cosmological stress-energy is an isotropic perfect fluid with energy density $\tilde{\rho} (t)$ and pressure $ \tilde{p} (t) $, which are related by an equation of state $\tilde{p} (t) = w \tilde{\rho} (t) $. We will show that in the AdS spacetime, the fluid is still isotropic on the cosmological slices with the same equation of state $p= w\rho $ and that the pressure in the AdS radial direction is given by $p_y= (3w - 1)\rho /2 $. The pressures and density decay towards the AdS boundary as well as in time as the universe expands.  A case of special interest is a free, massless scalar field which in a $(D-1)$-dimensional FRW spacetime has the equation of state $\tilde{p} = \tilde{\rho} $, that is, $w=1$. Hence the scalar field generates a $D$-dimensional AdS cosmology which is isotropic in all spatial directions and has corresponding equation of state $p=\rho$.  According to the AdS/CFT prescription, such a scalar field in the bulk AdS is dual to to a scalar operator in the CFT with vanishing expectation value but nonzero source.
    
We will focus on FRW metrics with negatively curved spatial slices, in which case~$\gamma_{ab}$ approaches the future Milne wedge of Minkowski space at late time, so long as $w>-1/3$.  The resulting AlAdS solution therefore approaches either the Poincar\'e patch of AdS or the AdS soliton at late times, which we interpret as an approach to equilibrium.  We use these solutions to perturbatively study the approach to equilibrium of the boundary stress tensor and the ADM charges. 
Interestingly, we find that the latter \textit{decrease} to their equilibrium values at late times, with the time dependent correction proportional to the dimensionless density parameter of the universe $\Omega$. 
For example, the mass of the solitonized cosmology decays as
\be
 \calm = \left( 1-{1\over 2} \Omega \right)  \calm^{(0)} + \cdots,
 \ee
 where $\calm^{(0)} $ is the mass of the static soliton and~$\cdots$ stands for subleading terms at late times.
In the context of spacetimes approaching the AdS soliton, this result is consistent with the energy conjecture of \cite{Horowitz:1998ha} that the AdS soliton is the lowest energy spacetime with the prescribed asymptotic structure.

A second application of our cosmological AdS solutions will be to compute the behavior of the entanglement entropy $S$ of a spherical region in the CFT as the spacetime evolves to equilibrium.  We use the covariant  prescription of~\cite{Hubeny:2007xt}, which is a generalization of the static prescription~\cite{Ryu:2006ef}.  This states that the entanglement entropy~$S_\mathcal{R}$ of a region~$\mathcal{R}$ of a holographic CFT is related to the area of a special bulk surface~$\Sigma$.  In general,~$S_\mathcal{R}$ is UV-divergent, but it can be regulated and the behavior of this regulated entropy~$S_\mathrm{ren}$ is studied.  We find that at late times,~$S_\mathrm{ren}$ decays as a power law in the proper time of an asymptotically static observer.

Our results add to and complement the substantial body of work in the literature on vacuum AlAdS spacetimes in which the metric on the AdS boundary is time dependent.  For instance,~\cite{Fischler:2013fba} constructed an elegant solution in which the metric on each Poincar\'e slice is a de Sitter cosmology.  Several studies in the general category of holographic cosmology apply coordinate transformations to AdS black holes to produce cosmological boundaries \cite{Lidsey:2009xz,Erdmenger:2012yh,Ghoroku:2012vi,Banerjee:2012dw,
Compere:2008us,Binetruy:1999hy,Kajantie:2008hh,Apostolopoulos:2008ru}, and resulting metrics have been analyzed as describing an expanding boost-invariant plasma \cite{Janik:2005zt,Janik:2006ft,Kajantie:2008jz,Culetu:2009xm,Pedraza:2014moa}.
Significant analytical work has also been done on out of equilibrium thermal properties of field theories using various AdS black hole spacetimes, including \cite{Janik:2010we,Lamprou:2011sa,Figueras:2009iu,
Tetradis:2009bk,Heller:2011ju,Fischetti:2012ps,Fischetti:2012vt,Beuf:2009cx}.  Discussion and further references can be found in \cite{Marolf:2013ioa}.  Our work adds a set of new 
non-vacuum AlAdS spacetimes which allow a wide range of boundary metrics.
  
This paper is organized as follows.  Section \ref{derive} contains the derivation of the new AlAdS solutions, as well as an analysis of the scalar field and perfect fluid cases.  In section \ref{sec:bst}, the leading time dependent corrections to the boundary stress tensor and the ADM charges for a solitonic cosmology in an open universe are found.  In section \ref{entropy} the perturbation to the entanglement entropy is calculated, and section \ref{conclusion} contains discussion and concluding remarks.  Unless otherwise specified, we take Newton's constant~$G_N = 1$.

\section{AdS and AdS soliton cosmologies}\label{derive}

We start by considering AlAdS spacetimes of the general form
\be
\label{metric}
ds_D^2 = dy^2 + e^{2y/l}\gamma_{\mu\nu } (x^\alpha ) dx^\mu dx^\nu,
\ee
where~$l$ is the AdS length and as before~$\mu,\nu = 0,1,\dots, D-1$.  For $\gamma_{\mu\nu} = \eta_{\mu\nu}$ this is AdS in Poincar\'e coordinates with cosmological constant given by 
\be
\Lambda = -{(D-1)(D-2)\over 2l^2},
\ee
so we will refer to hypersurfaces~$y = \mathrm{const.}$ as ``Poincar\'e slices''.  As mentioned above, it is well known that the Einstein equations with cosmological constant $\Lambda$ are still satisfied for any Ricci-flat $\gamma_{\mu\nu}(x^\rho)$.  In the particular case where~$\gamma_{\mu\nu}$ is a cosmological metric, we refer to \eqref{metric} as an AdS cosmology.  

In general, spacetimes of the form~\eqref{metric} with~$\gamma_{\mu\nu} \neq \eta_{\mu\nu}$ suffer from a singularity at the Poincar\'e horizon~$y \to -\infty$.  This singularity can be resolved by introducing an additional compact direction~$v$:
\be
\label{solmetric}
ds_D^2 = {dy^2 \over F(y) }  + e^{2y/l}\left( F(y) dv^2 + \gamma_{\mu\nu } (x^\rho ) dx^\mu dx^\nu  \right),
\ee
where $F(y) = 1- e^{ -(D-1)(y-y_+)/l }$, and now~$\mu,\nu= 0,1,\dots,D-2$.    The metric~\eqref{solmetric} is capped off at~$y = y_+$, so that the Poincar\'e horizon (and its possibly singular behavior) is removed.  Regularity at this cap fixes the period of~$v$ to be
\be\label{period}
v \sim v + \frac{4\pi l e^{-y_+/l}}{D-1}.
\ee
Now, if $\gamma_{\mu\nu } = \eta_{\mu\nu}$, then~\eqref{solmetric} is the usual AdS soliton metric~\cite{Horowitz:1998ha}\footnote{Although this is no longer Poincar\'e AdS, we will continue to call surfaces of~$y = \mathrm{const.}$,~$v = \mathrm{const.}$ Poincar\'e slices.}.  However, it was noted in \cite{Haehl:2012tw} that the Einstein equations with negative cosmological constant $\Lambda$ will still be satisfied for any Ricci-flat~$\gamma_{\mu\nu}$.  In analogy with~\eqref{metric}, if~$\gamma_{\mu\nu}$ is a cosmological metric, we will refer to~\eqref{solmetric} as an AdS soliton cosmology.

The solutions~\eqref{metric} and~\eqref{solmetric} provide a simple construction of AlAdS spacetimes with any desired Ricci-flat boundary metric~$\gamma_{\mu\nu}$\footnote{Technically, the boundary of~\eqref{solmetric} is~$\gamma_{\mu\nu}$ cross the circle direction~$v$.}.  Our goal is to generalize the above results to isotropic FRW cosmological metrics $\gamma_{\mu\nu}$; as such cosmologies are not (in general) Ricci-flat, we will require the introduction of matter fields.

\subsection{Massless Scalar Field}

We obtain a direct analogue of the results for vacuum metrics by considering a free massless scalar field $\phi$ in the AdS and AdS soliton spacetimes above.  The full $D$-dimensional Einstein-massless scalar equations are
\be
\label{einstein}
G_{ab}=-\Lambda g_{ab}+8\pi T_{ab},\qquad \nabla^2\phi=0,
\ee
where~$G_{ab}$ is the Einstein tensor and 
\be
T_{ab}={1\over 8\pi}[(\nabla_a\phi)\left(\nabla_b\phi\right)-{1\over 2}g_{ab}g^{cd}(\nabla_c\phi)\left(\nabla_d\phi\right)]
\ee
is the stress-energy of a free massless scalar field in any dimension.  
Consider a lower-dimensional metric $\gamma_{\mu\nu}(x^\rho)$ and scalar field configuration $\phi(x^\mu)$ that solve the Einstein-scalar equations
\be
\label{slice}
{\widehat G}_{\mu\nu} = 8\pi{\widehat T}_{\mu\nu}, \qquad \widehat\nabla^2\phi=0,
\ee
where hatted objects are computed with respect to the metric~$\gamma_{\mu\nu}$. Furthermore, let
\be\label{smetricdef}
s_{\mu\nu}= e^{2y/l} \gamma_{\mu\nu}
\ee
be the induced metric on a Poincar\'e slice. Finally, we pause to note that the scalar field stress energy satisfies the important property that from the full~$D$-dimensional point of view, the induced stress tensor on each Poincar\'e slice is equal to the lower-dimensional stress tensor of the scalar field on~$\gamma_{\mu\nu}$:
\be
\label{Tcondition}
T_{\mu\nu} = \widehat{T}_{\mu\nu}.
\ee

Now, consider first the metric (\ref{metric}).  The $D$-dimensional Ricci tensor for $g_{ab}$ is related to the Ricci tensor of $s_{\mu\nu}$ by
\be
\label{ricci}
R_{\mu\nu} [g] = R_{\mu\nu} [s] - \frac{D-1}{l^2} \, s_{\mu\nu}, \quad R_{yy} = \frac{D-1}{l^2}.
\ee
 When these components are assembled into the $D$-dimensional Einstein tensor and substituted into the left hand side of the Einstein field equation \eqref{einstein}, one sees that the terms which do not involve the curvature of $s_{\mu\nu}$ are equal to the cosmological constant term on the right hand side. If $\gamma_{\mu\nu}$ is Ricci flat, then the metric (\ref{metric}) is a solution with $T_{ab}=0$. If instead $\gamma_{\mu\nu}$ is a solution to (\ref{slice})
 with nonzero ${\widehat T}_{\mu\nu}$,
  then it is then straightforward to show that the metric~\eqref{metric} constructed from~$\gamma_{\mu\nu}$ will satisfy the full equations of motion \eqref{einstein}, with the full bulk scalar field taken to be~$\phi(x^\mu)$ (which, in particular, is independent of~$y$). The additional nonzero component of the
  stress-energy tensor is $8\pi T_{yy}=-{1\over 2} s^{\mu\nu} \nabla_\mu \phi \nabla_\nu \phi$. The construction with the solitonized metric~\eqref{solmetric} proceeds in a similar way, and one finds that $T_{yy}$ is the same and $T_{vv} = g_{vv} T_{yy}$.

One may naturally ask if such a straightforward foliation can be extended to other types of matter as well.  For instance, one might hope to replace the scalar field with a Maxwell field and obtain multi-black hole solutions analogous to those of~\cite{Kastor:1992nn}.  This is not the case: a key ingredient in the proof was the property \eqref{Tcondition} of the scalar field stress-energy. 
This property holds for the massless scalar field stress tensor but not {\it e.g.} for Maxwell fields, or even for a scalar field with nonzero potential $V(\phi)$.

A massless scalar field that depends only on time can serve as the source for an FRW cosmology on the Poincar\'e slices of either \eqref{metric} or the soliton metric \eqref{solmetric}.  For instance, setting $d\hat s^2=\gamma_{\mu\nu}dx^\mu dx^\nu$ and specializing to $4$-dimensional cosmologies with flat spatial sections we have
\be
d\hat s^2= -dt^2 + \left({t\over t_0}\right)^{2/3}(dx^2+dy^2+dz^2),\quad \phi=-\sqrt{{2\over 3}} \, \ln\left({t\over t_0}\right),
\ee
with corresponding stress tensor equal to that of a perfect fluid obeying the stiff matter equation of state $\tilde{p}=\tilde{\rho}$.  From the holographic perspective, the AdS/CFT dictionary tells us that the bulk scalar field is dual to a scalar operator in the CFT.  To be specific, the near-boundary behavior of a massless scalar field in AdS takes the form
\be
\phi(y) = \left(\phi_0 + \cdots\right) + e^{-(D-1)y/l} \left(\phi_{(D-1)} + \cdots \right),
\ee
where~$\phi_0$ and~$\phi_{(D-1)}$ are independent parameters that are fixed by the boundary conditions, and~$\cdots$ represent subleading terms in~$e^{-y/l}$.  The coefficient~$\phi_0$ should be interpreted as the source of a scalar operator~$\mathcal{O}$ of dimension~$D-1$, whose expectation value is~$\left\langle \mathcal{O} \right\rangle = \phi_{(D-1)}$.  Our solutions correspond to the special case~$\phi_{(D-1)} = 0$.  Note that this is unconventional: the operator~$\mathcal{O}$ is being sourced, but nevertheless has a zero expectation value.

\subsection{Perfect Fluid Matter}

Noting that property~\eqref{Tcondition} of the scalar field stress tensor was the key element in the above construction, we may extend the range of AdS and AdS soliton cosmologies by considering more general types of stress-energy that satisfy this condition.  We will shortly focus on perfect fluids, but we begin by assuming just  that the metric $\gamma_{\mu\nu}$ satisfies Einstein's equation on a Poincar\'e slice with some stress-energy ${\widehat T}_{\mu\nu}$.   We can then analyze the content of the full $D$ dimensional Einstein equations,
 beginning with the AdS type metrics (\ref{metric}), in the following way.  

Using the relations between the components of the Ricci tensor (\ref{ricci}) as in the previous subsection, we find that
 the AdS-type metric \eqref{metric} solves the Einstein equation~\eqref{einstein} with stress-energy given by
\be
\label{fullads}
T_{\mu\nu} = {\widehat T}_{\mu\nu},\qquad T_{yy}={1\over D-3}\, T ,
\ee
where $T=s^{\mu\nu}\,T_{\mu\nu}$. 
 A similar analysis for the AdS soliton-type metric \eqref{solmetric} shows that the Einstein equation \eqref{einstein} is solved with stress-energy given by
\be
\label{fullsoliton}
T_{\mu\nu} = {\widehat T}_{\mu\nu},\qquad T_{yy}={1\over D-4} \, g_{yy}T, \qquad T_{vv}={1\over D-4} \, g_{vv}T.
\ee
For example, one could embed a textbook example of a four-dimensional spherical static star into AdS. According to
(\ref{fullsoliton}) the pressures in the radial AdS and compact soliton directions of this cigar-star will be equal to each other,
but different from the radial pressure in the Poincar\'e plane.

We now specialize to the case of AdS and AdS soliton cosmologies, taking
the metric $\gamma_{\mu\nu}$ to have the FRW form
\be
\label{cosmometric}
d\hat s^2 =  -dt^2 + a^2 (t)\,  d\Sigma _k ^2,
\ee
where $d\Sigma _k ^2 $ is  a metric on a space with constant curvature $k=0,\pm 1$.  We also restrict our attention to $4$-dimensional Poincar\'e slices, so that the AdS cosmologies \eqref{metric} have overall dimension $D=5$ and the AdS soliton cosmologies \eqref{solmetric} have dimension $D=6$. 
 Finally, we assume that the stress-energy ${\widehat T}_{\mu\nu}$ on the slice has the perfect fluid form
\be
{\widehat T}_{\mu\nu} = (\hat\rho+\hat p)\hat{u}_\mu \hat{u}_\nu +\hat p \gamma_{\mu\nu},
\ee
with $\gamma^{\mu\nu}\hat{u}_\mu \hat{u}_\nu=-1$ and equation of state $\hat p=w\hat\rho$.  Note that the strong energy condition requires~$w \geq -1/3$.  Important special cases are $w=0$ for dust, $w=1/3$ for radiation, and $w=1$ for the massless free scalar field; indeed, note that such a stress tensor obeys the condition~\eqref{Tcondition}\footnote{Values of $w$ different from~1 could be obtained from an interacting scalar field, but as such interactions would require the introduction of a scalar potential, they are not compatible with our ansatz.  We will keep $w$ general, with the understanding that this is a bulk, hydrodynamic description.}.

The cosmological scale factor on the Poincar\'e slices evolves according to the Friedmann equations
\be
\label{frweqs}
d(\hat \rho\,  a^3 ) = - \hat p\,  d(a^3 ) , \qquad
\left( {\dot a\over a}\right)^2 ={8\pi \hat\rho\over 3}-{k\over a^2}.
\ee
The full stress energy tensor $T_{ab}$ for AdS cosmologies, given by \eqref{fullads}, now has the form of an anisotropic fluid, with a distinct equation of state parameter for the pressure in the $y$-direction.  Moreover, the energy density and pressures depend on the radial coordinate~$y$, as well as on time.  One finds that the energy density is given by $\rho= e^{-2y/l}\hat\rho$, while the pressures  tangent to the Poincar\'e slices satisfy an equation of state in the $D$-dimensions of the same form as in $(D-1)$, namely $p=w\rho$. In the $y$-direction one finds that $p_y = w_y\rho$ with
\be
w_y ={ (3w -1)\over 2}.
\ee
With a soliton, equation \eqref{fullsoliton} implies that the pressure in the compact $v$-direction is equal to $p_y$, so also
$p_v= w_y\rho$. To summarize, the stress-energy for the AdS soliton cosmology is
\be
\rho= e^{-2y/l}\hat\rho (t), \quad p=w\rho, \quad p_y =p_v = { (3w -1)\over 2} \,  \rho.
\ee
Some observations are as follows.  For $w=1$, which corresponds to the massless scalar field discussed above, $w_y=1$ as well so
the pressure in the full spacetime is isotropic.  For radiation ($w=1/3$), the stress-energy on the Poincar\'e slices is traceless and the pressure orthogonal to the slices vanishes, so the stress tensor remains traceless. For $w<1/3$, the orthogonal pressure is negative.

\subsection{Open AdS and AdS Soliton Cosmologies}

We will be particularly interested in AdS and AdS soliton cosmologies with open ($k=-1$) FRW universes on the Poincar\'e slices.  In this case, provided that the equation of state parameter is in the range $w>-1/3$ (that is, that the strong energy condition holds), the energy density $\hat\rho$ will fall off faster than $1/a^2$ and at late times the scale factor will grow linearly in time.  At sufficiently late times, the metric $\gamma_{\mu\nu}$ on the Poincare slices then approaches
\be
d\hat s^2_\mathrm{late} =  -dt^2 + t^2\,  d\Sigma _{-1} ^2,
\ee
which is flat spacetime in Milne coordinates.
The full AdS and AdS soliton cosmological metrics \eqref{metric} and \eqref{solmetric} then respectively approach the AdS or AdS soliton metrics at late times.  The late-time behavior of these cosmologies can therefore be thought of as an approach to equilibrium; in particular, the CFT dual can be thought of as an expanding isotropic plasma equilibrating at late time.  The solutions~\eqref{metric} and~\eqref{solmetric} correspond to the plasma being in a deconfined or confined phase, respectively.

\section{ADM Mass and Boundary Stress Tensor for AdS Soliton Cosmologies}
\label{sec:bst} 

From the field theoretic side, the late-time behavior of the AdS cosmology with open spatial slices described above is interpreted as a relaxation of the CFT to the vacuum state.  This relaxation can be studied by computing the late-time behavior of CFT observables. 
As  a first examination of the properties of these AdS cosmologies, we look at how the cosmological expansion impacts the boundary stress tensor and the
ADM mass and tensions of the AdS soliton (which corresponds to the confined phase of the dual field theory; see e.g.~\cite{Mateos:2007ay}).  In general we lack a definition of the ADM charges that will apply at the boundary $y=\infty$ with a time dependent boundary metric.  However, as we will see the special case of an open cosmology with matter obeying the strong energy condition~$w>-1/3$ allows for a perturbative computation of how the ADM charges of the soliton approach their static values at late times.

The static AdS soliton has negative ADM mass, reflecting the negative Casimir energy of the boundary field theory with a  compact direction, and is conjectured to be the lowest energy solution among spacetimes with these asymptotics \cite{Horowitz:1998ha}.
%
In addition to its mass, the AdS soliton has nonzero ADM tensions \cite{El-Menoufi:2013pza,El-Menoufi:2013tca}.  The tension along the compact $v$-direction in (\ref{solmetric})  is found to be large and positive, while the other three spatial tensions have negative values, such that the trace of the ADM charges (sum of the mass and the tensions) vanishes. One can think of the static soliton solution as an equilibrium configuration. In this section we will compute the approach to equilibrium of the boundary stress tensor, as well as the mass and tensions for an open AdS soliton cosmology. We will see that the mass decreases to the static soliton value, a result that is consistent with the minimum mass conjecture with matter obeying the strong energy condition.

As noted above, the FRW boundary metric does not have a time-translation symmetry and therefore the ADM mass is not defined in the usual sense.  However, at late times the FRW cosmologies with negatively curved  spatial slices approach Minkowski spacetime.  We can then define a time dependent ADM mass in this late time limit by writing the metric as static AdS plus time dependent perturbations that decay to zero. These perturbations to the metric determine the late time corrections to the asymptotic constant value of the mass of the soliton. 


Consider an AdS soliton cosmology \eqref{solmetric} with an open FRW metric 
\be
\label{openfrw}
d\hat s^2 = - dt^2 + a^2 (t) \left( d\chi ^2 + \sinh ^2 \chi \, d\Omega_{(2)} ^2 \right)
\ee
on the Poincar\'e slices.  We assume that $w>-1/3$, so that in the late time limit $a(t)\simeq t$.  Define new coordinates on the slices according to
\be
\label{coordtrans}
T =a(t ) \cosh \chi,  \quad R= a(t) \sinh \chi.
\ee
Note that since $R/T =\tanh \chi$ it follows that $R/T \leq 1$ with equality when $\chi \rightarrow \infty$.
In terms of these new coordinates the AdS soliton cosmology has the form
\be
\label{latemetric}
ds^2 = {dy^2 \over F(y) } + e^{2y/l} \left[\, F(y)\,  dv^2 -dT^2 (1-\delta \tilde{g}_{TT} ) +dR^2 (1+ \delta \tilde{g}_{RR} ) + 2 \, \delta \tilde{g}_{TR}\,  dR\, dT  +
R^2 d\Omega_{(2)}  ^2 \right]
\ee
where 
$F(y)$ is given in (\ref{solmetric}) and the functions $\delta \tilde{g}_{TT}$, $\delta \tilde{g}_{RR}$ and $\delta \tilde{g}_{TR}$, which give the deviance of the metric on the Poincar\'e slices from flat, may be written as
%
%
%
\be
\label{fndef}
\delta \tilde{g}_{TT} = \Omega \, \frac{1}{1-(R/T)^2}, \quad
 \delta \tilde{g}_{RR} =  \Omega \, \frac{(R/T)^2}{1-(R/T)^2}, \quad \delta \tilde{g}_{TR} = \Omega \, \frac{R/T}{1-(R/T)^2}.
\ee
Here $\Omega$ is the dimensionless density parameter of the open FRW metric,
\be\label{omegadef}
\Omega = {8\pi \hat \rho \over 3 H^2} = \left(1-  {1\over \dot{a} ^2 }  \right),
\ee
and $H=\dot a/a$ is the Hubble parameter. For an open universe $\Omega<1$ and approaches zero in the far future. Hence, the metric \eqref{latemetric} approaches the AdS soliton at late times.  We emphasize that the expressions in \eqref{fndef} are exact up to this point.

To proceed further, the density parameter  $\Omega$  must be expressed in terms of the asymptotically Minkowski coordinates $(T,R)$, which requires the expression for the scale factor $a(t)$ at late times.  To obtain this expression, first we substitute the equation of state~$p=w \rho$ into the Friedman equations \eqref{frweqs}, which allows the energy density to be solved for in terms of the scale factor, giving
%
\be\label{laterho}
\hat\rho (t)=  {3\bar{\Omega}_* H_*^2\over 8\pi (H_*a(t))^{3(1+w)}} \  ,  \mbox{ where }  \bar{\Omega}_*  \equiv {\Omega_* \over (1- \Omega_*  )^{3(w+1)/2 }}
\ee
and $H_*$ and $\Omega_*$ are the Hubble and density parameters evaluated at a fiducial time $t=t_*$.
The density and scale factor evaluated at $t_*$ are given by  $\hat{ \rho}_* = 3\Omega_* H_*^2/8\pi$ and  $a_* = 1/ ( H_*  \sqrt{1-\Omega_* }) $ respectively.   The equation for the scale factor then reduces to
\be
\dot a^2 = 1 +{ \bar\Omega_*H_*^2 a_*^{3(1+w)}\over a^{1+3w}}.
\ee
For $w>-1/3$, this reduces in the limit of large scale factor to $\dot a^2 \simeq1$, giving $a(t)\simeq t$ in the late time limit.  Including a subleading correction of the form~$a(t) \simeq t + \alpha t^\beta$ yields
\begin{subequations}
\label{latea}
\bea
a(t)  &\simeq t - {\bar\Omega_* \over  6wH_* ( H_* t ) ^{3w } }, \quad  w\neq 0, \\
a(t) &\simeq t + {\bar\Omega_* \over 2 H_*  } \ln \left( {t \over H_* } \right) ,\quad  w = 0,
%
\eea
\end{subequations}
which by~\eqref{omegadef} yield
\be\label{omegat}
\Omega (t)  \simeq {\bar\Omega_*\over  (H_* t) ^{(1+3w )}}.
\ee

The expressions~\eqref{latea} can be inverted and combined with the transformation to~$(R,T)$ coordinates to yield the coordinate transformation from~$t$ to~$(R,T)$, valid at late times, including terms up to order $R^2 / T^2$,
\begin{subequations}
\bea
t&\simeq T+{\bar\Omega_*\over 6wH_*(H_*T)^{3w}}-{R^2\over 2T},\qquad w\neq 0, \\
t&\simeq T -{\bar\Omega_*\over 2H_*} \ln\left( {T \over H_* } \right)   -{R^2\over 2T},\qquad w= 0,
\eea
\end{subequations}
which then give $\Omega$ as a function of $T$ and $R$:
\begin{subequations}
\label{lateomega}
\bea
\Omega &\simeq  {\bar\Omega_*\over ( H_* T) ^{3w+1 } } 
\left(1- {(1+3w)\bar\Omega_*\over 6w (H_* T ) ^{3w +1 } }  + {(1+3w ) R^2 \over 2T^2  } \right), \quad  w\neq 0, \\
\Omega &\simeq  {\bar\Omega_*\over H_* T} \left( 1  +
  {\hat\Omega_* \over 2 H_* T } \ln\left( {T \over H_* } \right)    +{ R^2 \over 2 T^2} \right), \quad  w = 0.
\eea
\end{subequations}
This is our desired result.

We are now prepared to compute the leading late time corrections to the boundary stress tensor density, which we will denote by
 $\tau _{\mu\nu}$. Let $K _{\mu\nu}$ be the extrinsic curvature of the  AdS boundary.  In the boundary stress tensor formalism a boundary
 action is defined that includes an integral over $K$ plus
 geometrical counterterms that are constructed from the metric on the boundary $s_{\mu\nu}$, defined in equation \eqref{smetricdef}.
  These terms include a cosmological constant,
  the scalar curvature of $s_{\mu\nu}$, and potentially higher derivative counter terms as needed. 
 The stress tensor density  results from 
varying the boundary action with respect to $s^{\mu\nu}$. The coefficients of the counter terms are
 chosen to cancel divergences that occur in $\tau_{\mu\nu}$ and are dimension dependent. One finds the result \cite{Balasubramanian:1999re,Myers:1999psa,de Haro:2000xn}
\be
\label{bst}
8\pi  \tau_{\mu\nu} = \sqrt{-s} \left( K _{\mu\nu} - Ks_{\mu\nu}  +{ D-2 \over l} s_{\mu\nu}  +{1\over D-3} G _{\mu\nu}  [s] + \cdots \right),
\ee
where the $\cdots$ indicate higher derivative terms in the Riemann tensor of $s_{\mu\nu}$, which we will show are subdominant at late times.
We work with the boundary stress tensor density because the volume element of the late time metric changes at leading order, and 
also because this is the appropriate quantity to integrate to get the ADM charges.

 In the case of the static AdS soliton, the metric on the boundary is flat and the terms in (\ref{bst})
  depending on the curvature of $s_{\mu\nu}$ all vanish.  This is no longer true for
 the cosmological AdS spacetimes. 
 The Einstein tensor  term in (\ref{bst}) 
 contributes a time-dependent piece to $\tau_{\mu\nu}$ which goes to zero at late times like the energy density $\hat\rho$ in (\ref{laterho}).  Additional time dependence
 in $\tau_{\mu\nu}$ comes from the volume element in (\ref{bst}) which goes like
 \be\label{volchange}
 \sqrt{-s} = \left(1-{1\over 2} \, \Omega \right) \sqrt{-s_{(0)} },
\ee
where $\sqrt{-s_{(0)} }$ denotes the volume element in the static AdS soliton, and the late time behavior of the density parameter
$\Omega$ is given in (\ref{omegat}).
Comparing the decay rates of $\hat\rho$ and $\Omega$, one finds that the 
contribution of  $G_{\mu\nu}$ to the boundary stress tensor density is subdominant at late times compared to that of the volume element.  The contributions of higher derivative terms in (\ref{bst}) will decay
 even more rapidly. The  leading contributions to the boundary stress tensor density are then readily found by combining
  the results for the static AdS soliton in \cite{El-Menoufi:2013pza} with  equation \eqref{volchange} giving
  %
  \be\label{solitonbst}
  \tau_{\mu\nu  } 
  = {e^{5y_+/l} \over 16\pi   l} \left( 1-{1\over 2} \, \Omega \right) \mathrm{diag} (-1, 1 ,1, 1, -4),
  \ee
  where the coordinates are ordered according to $(t, x_1, x_2 , x_3 , v)$. 
Hence the decaying time dependent corrections to the static values of $\tau_{\mu\nu  } $ are simply proportional to $\Omega$, the density parameter of the cosmology.

We now use the results above to determine the ADM mass and tensions for the AdS soliton cosmologies.  Comparison with \cite{El-Menoufi:2013pza} shows that the integrands of the ADM charges in AdS coincide with the first three terms in the boundary stress tensor in equation (\ref{bst}), and the components of  $\tau_{\mu\nu} $ above then just need to be integrated to obtain the ADM charges. In the static coordinates, the density parameter $\Omega$ depends on $R$ as well as $T$, so the integrand is not a constant. This does not mean that  $R=0$ is a special point, since any location in the homogeneous open cosmology could equally well be chosen as the origin. For the static AdS soliton, the ADM charges are made finite by taking the planar geometry to be periodically identified, with {\ $-L_j /2 \leq x^j \leq L_j /2$. For notational brevity let the asymptotic volume be $V= L_1 L_2 L_3 L_v$ where  $L_v$ is the range of compact coordinate $v$ given in (\ref{period}). In the limit that the plane is infinite, the relevant energy is the mass per unit volume obtained by dividing the total mass by $V$, and similarly for the spatial tensions.
  
Finally, it is important to note that the static radial coordinate $R$ has the range $0\leq R\leq T$, with the upper limit corresponding to $\chi \rightarrow \infty$ in the coordinate transformation (\ref{coordtrans}).
The integrals for the ADM charges are then over a box of length $L< T$, and at the end we divide out the volume of the box.
Define the spatial average of the density parameter $\Omega$ at time $T$ by
\be
\label{avomega}
\ev{\Omega} 
= {\bar\Omega_*\over  V  ( H_* T) ^{3w+1 }  } \int   dx_1 dx_2 dx_3 dv \, \Omega (R , T).
\ee
In the late time limit, we substitute the approximate expression for~$\Omega$ given in (\ref{lateomega}).
For the general case $w\neq 0$, this yields
\be\label{intomega}
\ev{\Omega}  = {\bar\Omega_*\over V ( H_* T) ^{3w+1 }  } \left(1- {c_1\over T  ^{3w +1 } } + { c_2L^2 \over T^2 }\right),
\ee 
where the coefficients $c_1 , c_2$ can be read off of the expansion of $\Omega$ in equation (\ref{lateomega}). One sees that
the terms proportional to $c_1$ and $c_2$ make increasingly small contributions and so will be dropped in subsequent formulae.  This also allows us to treat the cases~$w = 0$,~$w \neq 0$ simultaneously, since the leading-order term in~\eqref{intomega} is identical to that obtained in the special case
 $w=0$.

Following the conventions of past work ({\it e.g.} \cite{El-Menoufi:2013pza}), we give the ADM tension rather than a pressure, where tension is simply minus the pressure\footnote{This convention
is natural in asymptotically flat static spacetimes where the gravitational tension can be shown to be positive \cite{Traschen:2003jm}.}.
 Assembling the pieces, at late times the mass and tensions of the soliton in the metric \eqref{latemetric} are
\begin{subequations}
\label{admcharges}
\bea
\calm =  \calt _j &  = -{ V  \over 16 \pi  l } e^{5y_+/l}\left( 1 -{1\over 2}\ev{\Omega}  \right) \  , \quad j=1,2,3, \\
\calt_v &=   {4V  \over  16 \pi l }e^{5y_+/l}  \left( 1 -{1\over 2} \ev{\Omega}  \right).
\eea
\end{subequations}
The expressions for the ADM charges have the same structure as the components of the boundary stress tensor, relaxing to the 
equilibrium values like $\ev{\Omega} $. 
Since~$\ev{\Omega} > 0$, the mass of the AdS soliton cosmology decreases as $ \ev{\Omega} $ goes to zero, approaching 
its negative static value at late times, consistent with the energy bound conjectured in \cite{Horowitz:1998ha}.  The tension $\calt_v$ around the compact dimension increases to its static positive value, while the trace $\calm + \calt_v +\Sigma _j \calt_j $ vanishes
throughout the relaxation process.

\section{Entanglement Entropy}
\label{entropy}

The new AdS cosmological solutions allow us to  compute how the entanglement entropy of a region in the dual CFT  approaches equilibrium.  To perform the computation, we use the holographic prescription \cite{Ryu:2006ef,Hubeny:2007xt}, which proposes that the entanglement entropy of a region~$\mathcal{R}$ (called the entangling region) in the boundary CFT is equal to
\be
\label{sa}
S_\mathcal{R} = \frac{\mathrm{Area}\left[\Sigma\right]}{4G_N},
\ee
where~$\Sigma$ (referred to as the entangling surface) is the minimal-area extremal surface in the bulk spacetime anchored to~$\partial\mathcal{R}$ and homologous to~$\mathcal{R}$. Note that in this section we have restored Newton's constant $G_N$.  We will also keep~$w$ general, though we emphasize that only the case~$w = 1$ (wherein the bulk matter is a scalar field) has a well-understood CFT dual.

Parametrizing~$\Sigma$ as~$X^a(\sigma^i)$, with~$\sigma^i$ coordinates on~$\Sigma$, $i=1,..., D-2$, the area functional is
\be
\label{eq:area}
A = \int \sqrt{h} \, d^{D-2} \sigma,
\ee
where~$h$ is the determinant of the induced metric on the surface
\be
\label{eq:inducedh}
h_{ij} = g_{ab} \partial_i X^a \partial_j X^b.
\ee

In general, extremizing~\eqref{eq:area} to obtain the entangling surface is difficult to accomplish analytically, and the AdS soliton cosmologies
are no exception.  However, we can make progress  by working in the non-solitonized AdS cosmology~\eqref{metric} and noting that the calculations performed there should approximate those in the AdS soliton cosmology, as long as the relevant surfaces do not extend too deeply into the spacetime.  The boundary metric is then
\be
\label{eq:bndryflat}
ds^2_\partial = -dT^2 (1-\delta \tilde{g}_{TT}) + dR^2 (1+\delta \tilde{g}_{RR}) - 2\delta \tilde{g}_{TR} \, dR \, dT + R^2 d\Omega_{(2)}^2,
\ee
and the full metric is given in (\ref{metric}) with $\gamma_{\mu\nu}$ equal to $ds^2_\partial$.
Working in pure AdS has the significant advantage that the extremal surface is known for a spherical entangling region on the boundary \cite{Ryu:2006ef}. This allows us to use perturbative techniques to compute the time dependent correction to the area as the metric approaches the static AdS spacetime in the future.

In order to compute the late-time behavior of the entanglement entropy we work to first order  in powers of $R/T$ in~$\delta \tilde{g}_{TT}$,~$\delta \tilde{g}_{RR}$, and~$\delta \tilde{g}_{TR}$, given in equations \eqref{fndef} and \eqref{lateomega}.  We take the boundary of the entangling region to be a sphere of radius~$R_0$ at some time~$T_0$; the corresponding entangling surface~$\Sigma$ in pure AdS was found in \cite{Ryu:2006ef}.  We may then perturb off of this solution to compute the leading correction to the area.  There are two natural options for how this sphere should evolve in time: (i) the sphere can be of fixed proper size in the asymptotically static coordinates so that~$R_0$ is held constant as~$T_0$ advances; or (ii) the sphere can be comoving, so
that fluid elements on the boundary of the sphere follow geodesics, and~$R_0$ grows like~$a(t)$.  We will discuss both choices below.

\subsection{Zeroth Order Solutions}

At zeroth order, the boundary metric~\eqref{eq:bndryflat} is just Minkowski space.  Parametrizing the surface by~$z \equiv l e^{-y/l}$ and the coordinates on the sphere, the area functional~\eqref{eq:area} is
\be
A  = 4\pi l^3 \int _\epsilon ^1 dx \, {(1-x^2 )^{1/2} \over x^3 },
\ee
where $\epsilon = z_\mathrm{cut} / R_0 $ and $z_\mathrm{cut}$ is a UV cutoff to regulate the integral.  The corresponding entangling surfaces were calculated in~\cite{Ryu:2006ef} and are given by 
\be
\label{zerosurf}
\Sigma_{0}: \quad z^2 + R^2 = R_0 ^2  \  , \quad T = T_0,
\ee
with area 
\be
\label{zeroarea}
A^{(0)} = l^3 \left[ {A_\mathrm{static} \over  2 z_\mathrm{cut} ^2} - \pi \ln\left( \frac{A_\mathrm{static} }{ \pi z_\mathrm{cut}^2 }\right)-\pi \right],
\ee
where~$A_\mathrm{static} = 4\pi R_0 ^2$ is the area of~$\partial\Sigma _0 $.  The first term in the above expression denotes the usual area law growth of the entanglement entropy, while the coefficient of the logarithmically divergent term provides a UV-independent measure of the entanglement entropy.

\subsection{First Order Corrections: Approach to Equilibrium}

Now, consider corrections to~\eqref{zeroarea} which arise both from perturbations to the metric and to the surface~$X^a$. Write each as a zeroth order piece plus a perturbation,
\be
\label{realdeal}
g_{ab} = g_{ab}^{(0)} + \delta g_{ab} \  , \quad \  X^a (\sigma_i ) = X^a _{(0)} + \delta X^a .
\ee
To first order  the volume element on the surface becomes
\be
\label{deltaa}
h = h^{(0)} \left( 1 + \Tr\left[ \delta g_{ab}\partial_i X_{(0)}^a \partial_j X_{(0)}^b + 2 g_{ab} ^{(0)} \partial_i \delta X^a \partial_j X_{(0)}^b\right]\right).
\ee
However, the second term in the trace is a variation of the surface in the background metric, and so this integrates to zero since the background surface is extremal. Thus the first-order change in the area is governed by the perturbation to the metric:
\be
\label{deltaatwo}
\delta A  ={1\over 2} \int \sqrt{h^{(0)} } \, \Tr\left[ \delta g_{ab}\partial_i X_{(0)}^a \partial_j X_{(0)}^b\right] \, d^{d-1} \sigma.
\ee
The final step is to substitute the expressions for the metric perturbations (\ref{fndef}) into the metric equations (\ref{metric}), (\ref{eq:bndryflat}).
 Using $R^\prime (z) = -z/R $ on
the zeroth order surface \eqref{zerosurf}, the induced metric in the perturbed spacetime is given by
\be
\label{latethree}
(g_{ab} ^{(0)} +\delta g_{ab} ) \partial_i X_{(0)}^a \partial_j X_{(0)}^b d\sigma^i d\sigma^j|_{\Sigma_0} 
= {l^2   \over z^2 }  \left\{  \left(  {R_0^2 \over R^2 } +  {z^2\Omega \over  T_0^2}  \right)  dz^2
+ R^2 d\Omega_{(2)} ^2 \right\},
\ee
where  $R=\sqrt{ R_0 ^2 -z^2 }$. Using this expression in (\ref{deltaatwo}) and substituting $\Omega$ from (\ref{lateomega}) gives
 the first-order correction to the area of the entangling surface
\bea
\label{latearea}
\delta A &=    { 4\pi l^3 \obar\over  ( H_* T_0 ) ^{3w+1} } \left( { R_0 ^2 \over  2T_0^2} \right) \int _\epsilon ^1 dx {(1-x^2 )^{3/2} \over x } \\
		 &=  { l^3\obar \over 4 (H_* T_0 ) ^{3w+3}}  H_*^2 A_\mathrm{static}   \left( \ln\left( {A_\mathrm{static}\over \pi  z_\mathrm{cut} ^2}\right) -{4\over 3} \right).
\eea
This result is valid at sufficiently late times such that $H_* T_0 \gg 1$ and   $T_0 \gg R_0 $. 

The entanglement entropy, including the leading late time contribution, follows from substituting
(\ref{latearea}) and (\ref{zeroarea}) into the entropy-area relation in equation (\ref{sa}). 
  The conversion from area to entropy contains the prefactor $l^3/G^{(5)}_N$, which can be translated into the parameters of the dual CFT.  According to the AdS/CFT correspondence, the solutions~\eqref{metric} in~$D = 5$ are dual to an~$\mathcal{N} = 4$ supersymmetric Yang-Mills theory on the FRW spacetime \eqref{cosmometric}.   Following  the discussion in \cite{Ryu:2006ef}, we consider ${\cal N} =4 \ SU(N)$ SYM theory on AdS$_5 \times S^5$, 
  in which case the AdS radius, the ten dimensional Newton's constant, and  the five dimensional Newton's constant are identified with the string coupling, string tension, and $N$ according to $l^4 = 4\pi g_s  (\alpha^\prime)^2 N$, $G^{(10)}_N =  8\pi^6 g_s^2  (\alpha^\prime)^4 $, and  
$G^{(5)}_N =G_N^{(10)} / l^5$.  This gives~$l^3/G^{(5)}_N = 2N^2/\pi $, and the entanglement entropy is then
\be\label{eq:Stot}
S = \frac{N^2 }{2\pi }\left[\frac{A_\mathrm{static}}{2z_\mathrm{cut}^2} - 
\pi  \ln  \left( {A_\mathrm{static}\over \pi  z_\mathrm{cut} ^2}\right) -\pi
+  \frac{ \obar H_*^2  A_\mathrm{static}}{4( H_* T )^{3w+3}   }\left(   \ln  \left( { 2A_\mathrm{static}\over \pi  z_\mathrm{cut} ^2}\right) - {4\over 3}       \right) + \cdots
\right],
\ee
where $\cdots$ denotes terms that are subleading at late time,  and the subscript on $T$ has been dropped for simplicity. The coefficient of the logarithmic term is invariant under rescalings of the cutoff, so it serves as a regularized measure~$S_\mathrm{ren}$ of the entanglement entropy.  One finds
\be
\label{eq:Sren}
\delta S_\mathrm{ren} \simeq \frac{N^2 }{8\pi} \frac{ \obar H_*^2  A_\mathrm{static}}{( H_* T )^{3w+3}}.
\ee
Note that this is positive, which means that~$S$ \textit{decreases} to its equilibrium value.  This behavior differs markedly from that of quenches in CFTs~\cite{Calabrese:2004eu,AbajoArrastia:2010yt,Albash:2010mv,
Balasubramanian:2010ce,Caceres:2012em,
Alishahiha:2014cwa,Alishahiha:2014jxa}, wherein the entanglement entropy grows until is saturates.  There is a temptingly simple and compelling physical reason for the decrease of $S$ found in this calculation: in the Cartesian coordinates~$T,R$ there is a nonzero radial flux proportional to $g_{TR}$, so the decrease of the entropy in the ball $R \leq R_0$ can be interpreted as due to an energy flow out of the ball.  In particular, in the quasi-particle picture of entanglement entropy propagation~\cite{Calabrese:2004eu}, entanglement is carried by entangled particle pairs; a new flow of such particles out of the entangling ball~$R \leq R_0$ leading to a decrease in entanglement entropy is consistent with this picture.  Alternatively, note that at late time, our bulk solution approach the Minkowski vacuum, and therefore the CFT evolves from an excited state to the zero-temperature vacuum state.  We would therefore naturally expect probes of correlation (such as entanglement entropy) to decay in the late time limit\footnote{We thank Juan Pedraza for this observation.}.

The time dependent contribution decays as a power law, and the time scale for the decay is set by the Hubble parameter $H_*$. The power  depends on the equation of state. For example, for dust the correction goes to zero like $T^{-3} $, and for a free massless scalar field like $T^{-6} $. The time dependence in $\delta S$ is analogous to the result of \cite{Engelhardt:2013jda}, in which the entropy of a strip in a vacuum-Kasner AdS spacetime was found to have a power law behavior, in both cases a reflection of the time evolution of the cosmology.

Turning to the amplitude of $\delta S_\mathrm{ren}$, we see that this is set by an interesting combination of factors. At sufficiently late times $t_* \Omega_* \ll 1$, so that $H_* ^2  \obar \simeq 8 \pi G_N ^{(5)}\hat\rho_* / 3 $.
Hence the dimensionless combination $H_* ^2  \obar A_\mathrm{static}$  has the interpretation of the non-vacuum energy, measured in Planck units, that is
contained in a shell of width the Planck length that surrounds the sphere. That is, the entangling modes of the perturbation 
act like they are concentrated on the boundary of the sphere. This is a reflection of the fact that the change in the area of the 
 extremal surface comes from the metric perturbations near the surface.

\subsection{The Cosmological View}

An alternative way to interpret the time evolution of the entanglement entropy is to take the boundary sphere to be comoving, so that
points on the boundary sphere follow geodesics. 
 In the cosmological coordinates \eqref{openfrw} this means that  the sphere is at a fixed coordinate $\chi= \chi _0$.  The extremal surface $\Sigma_0$ does not lie within a slice of constant cosmological time, but it does intersect the boundary at a constant time, as can be seen by 
evaluating  \eqref{coordtrans} at $z=0$.  
Transforming the zeroth order surface (\ref{zerosurf}) to  the cosmological coordinates gives
 \be\label{cosmosurf}
 \Sigma_0 : \quad   a(t) \cosh \chi = a( t_b )  \cosh \chi_0 \  , \quad z^2   + \cosh^2 \chi_0   \tanh ^2 \chi  = a(t_b )^2 \sinh ^2 \chi _0.
 \ee
Let
\be
A_\mathrm{geod}(t_b )  = 4\pi a( t_b )^2 \sinh ^2 \chi _0
\ee
be the proper area of the comoving  sphere on the boundary at $t_b$.
 Then in terms of the cosmological coordinates the zeroth order area (\ref{zeroarea}) becomes
\be\label{areacosmo}
A^{(0)} (t_b ) =   l^3 \left[ { A_\mathrm{geod}\over 2  z_\mathrm{cut} ^2}  - \pi \ln \left({  A_\mathrm{geod} \over \pi  z_\mathrm{cut}^2 } \right) -\pi 
 \right] 
\ee
and the time dependent correction (\ref{latearea}) is
\be\label{dacosmo}
\delta A = l^3  \obar H_*^2 (4\pi a_*^2 \sinh ^2 \chi _0 ) \left(  {a_* \over  a(t_b ) }   \right)^{3w+1}    
 \left( \ln\left( { A_\mathrm{geod} \over 2\pi z_\mathrm{cut}^2 }\right) -{2\over 3} \right).
\ee
Hence the time dependent piece redshifts to zero as 
 $(1+z_\mathrm{b} )^{-(3w+1)} $ where  $1+z_b = a(t_b ) / a_* $ is the cosmological redshift.
The power in the decaying term is different than for the static sphere (\ref{latearea}) because $A_\mathrm{geod} $ increases
 as $a^2$. 
 
 So far the expressions for the area (\ref{areacosmo}) and (\ref{dacosmo}) are just translations from
the asymptotically static coordinates to the cosmological coordinates. 
The difference 
from the previous section, in which the area of the boundary sphere is held constant,
 comes when one follows the time evolution by considering increasing values of the boundary time $t_b$.
The area of the boundary co-moving sphere increases like $a^2 (t_b )$, so  
 although (the UV-independent part of) $\delta A$ is positive and decreasing to zero, the total entropy increases with $t_b$. This brings up the important issue of 
 the range of validity of the expressions in cosmological time. As discussed in
 section \ref{entropy}, if the results are to be good approximations to the results in a solitonized spacetime, one needs to
 restrict to surfaces that do not penetrate too deeply into the bulk, which precludes taking~$R \sim t_b \sinh \chi _0$ too large. This means
 that the validity of (\ref{latearea}) is restricted to times that are not so large that the proper radius of the boundary sphere
 approaches the length scale set by the soliton, that is, we need $ t_b \sinh \chi _0 \ll l e^{-y_+ /l}$. The situation here
  is similar to that in \cite{Engelhardt:2014mea}.

\section{Discussion}
\label{conclusion}

In this paper, we have shown how to construct AdS cosmologies that satisfy the Einstein equations with a nonzero stress tensor and negative cosmological constant.  Our solutions were built as foliations of lower-dimensional solutions; the induced metric on each of these hypersurfaces itself satisfies Einstein's equations with a nonzero stress tensor.  This construction has the advantage that the boundary metric of the AlAdS solution is just (conformal to) the induced metric on each hypersurface.  Therefore, this construction offers us significant freedom in constructing AlAdS spacetimes with a boundary metric of our choosing.

The particular AdS cosmologies that we have constructed take the stress tensor to be that of a perfect fluid obeying the strong energy condition; for the equation of state~$w = 1$, the fluid is sourced by a massless noninteracting scalar field.  Moreover, we have focused on the specific case in which the spatial slices of the FRW cosmologies are negatively curved, as such FRW cosmologies then approach the Milne patch of Minkowski space at late times.  The AdS cosmology constructed from these slices therefore approaches the Poincar\'e patch of AdS at late times, while the AdS soliton cosmology approaches the static AdS soliton.

Such solutions are especially interesting because they allow us to perturbatively calculate the behavior of physically relevant quantities at late times.  For instance, we have calculated the late-time perturbation to the ADM mass of the AdS soliton cosmology, and have found this perturbation to  be
\be
\label{deltam}
\delta\calm  =  -{ \Omega \calm\over 2},
\ee
where~$\calm$ is the unperturbed mass and~$\Omega$ is the dimensionless density parameter of the FRW cosmology, which goes to zero at late times in the solutions we are interested in.  Since $\calm$ is negative for the soliton this implies that the mass \textit{decreases} to the mass of the static AdS soliton.  Hence this result is consistent with the energy conjecture of \cite{Horowitz:1998ha} that the AdS soliton is the lowest energy spacetime with the prescribed asymptotic structure.  We found the ADM tensions to be modified in a similar manner to~$\calm$.

Moreover, our solutions also have immediate applicability to large-$N$, strongly coupled CFTs via the AdS/CFT correspondence.  Indeed, our AdS cosmologies are dual to CFTs living on an FRW cosmology, while the massless, noninteracting scalar field in the bulk is dual to a scalar operator in the CFT with zero expectation value but nonzero source.  This atypical behavior can be tied to the fact that our AdS cosmologies are singular at the Poincar\'e horizon.  This singularity is removed by ``solitonizing'': introducing a compactified direction in the bulk that caps off the geometry.  This cap amounts to putting the CFT in a confined phase. 

As a probe of the behavior of the CFT on these FRW spacetimes, we study the entanglement entropy~$S$ of a sphere of constant radius.  Using perturbative techniques to find the leading time dependent correction to the entropy, we find that the regulated entanglement entropy $S_\mathrm{ren}$ decays as a power law to its equilibrium value.  The power depends on the equation of state of the fluid.  Note that this decay to equilibrium is starkly different from the behavior of entanglement entropy after a quench, when the entanglement entropy \textit{grows} to its equilibrium (thermal) value.  However, note that our solutions at late time approach Poincar\'e AdS, which has zero temperature; this is drastically different from the end state of a quench, which in the bulk is usually modelled by the injection of energy, forming a black hole of finite temperature.

Several issues and questions are raised by these examples.  First, can these solutions be generalized to include planar black holes in the bulk spacetimes?  Such solutions could conceivably be used to model the approach of a CFT on an FRW cosmology to thermal equilibrium with a nonzero temperature~$T$, much as here we have modeled the approach to equilibrium at~$T = 0$.
Second, for reasons of tractability our entropy calculations have been perturbative analyses in the AdS cosmology.  It would also be interesting to study the entanglement entropy of the CFT at all times, in both the AdS cosmologies and especially in the AdS soliton cosmologies, to see if there is any behavior that was not captured by our perturbative methods.  Such calculations would most likely be numerical, so we leave them for the future.

\section*{Acknowledgements}

The authors with to thank Netta Engelhardt and Don Marolf for useful discussions.  We also thank Juan Pedraza for useful comments on an earlier version of this paper.  This project was supported in part by the National Science Foundation under Grant No PHY11-25915, by FQXi grant FRP3-1338, and by funds from the University of California.

\bibliographystyle{JHEP}
\providecommand{\href}[2]{#2}\begingroup\raggedright\endgroup
\end{document}